\documentclass[12pt]{article}

\usepackage{amsmath, cite, graphicx, subfigure, hyperref}

 \numberwithin{equation}{section}

\usepackage{graphicx, subfigure}

\begin{document}

\title{\Large\bf Self-consistent field theory for the\\
distal ordering of adsorbed polymer:\\
comparison with the Scheutjens--Fleer model}

\author{Karl Isak Skau, Edgar M. Blokhuis, and Jan van Male\\
\textit{\normalsize Colloid and Interface Science, Leiden Institute of
Chemistry,}\\ \textit{\normalsize Gorlaeus Laboratories, P.O. Box
9502, 2300 RA Leiden, the Netherlands.}}  \maketitle



\noindent
\textbf{Abstract.} In a previous article [E.M.~Blokhuis, K.I.~Skau,
and J.B.~Avalos, J.~Chem.~Phys.~{\bf 119}, 3483 (2003)], a
self-consistent field formalism was derived for weakly adsorbing
polymers, valid for any chain length. It was shown that the presence
of a solid wall induces an ordering of the polymers on the scale of
the radius of gyration far away from the surface (the distal
region). These oscillations in the polymer concentration profile were
first noted in work by Semenov \emph{et al.}, and later observed in
numerical solutions of the Scheutjens and Fleer self-consistent field
model. In the present paper, we compare the weak adsorption model in
more detail with the numerical results from the Scheutjens and Fleer
model. Quantitative agreement is obtained for the polymer segment
density profile in a good (athermal) solvent. For theta solvent and
poor solvent conditions, it is necessary to extend the weak adsorption
model to take the non-local character of the polymer--solvent
interaction into account. Again, quantitative agreement is obtained
for the polymer segment density profile, in particular for the
transition from an oscillatory decaying profile to a monotonically
decaying profile when the bulk polymer density is below a certain
threshold value.


\pagebreak

\section{Introduction}

The adsorption of polymer onto a solid surface has received
considerable attention both from an experimental and a theoretical
point of view. Interest in these systems is driven by practical
applications, but also as a testing ground for various theoretical
approaches \cite{de_Gennes_book, Fleer_book, Eisenriegler_book}.

Theoretically, surfaces with enhanced polymer adsorption were first
studied by de Gennes \cite{de_Gennes_book, de_Gennes} in the context
of the Edwards self-consistent field theory \cite{Edwards}.  In the de
Gennes free energy functional, the so-called \emph{ground state
dominance approximation} \cite{de_Gennes_book, Edwards, Lifshitz} is
made in which the polymer chain length is essentially set to
infinity. Various extensions of the de Gennes model have been
formulated. In work by de Gennes \cite{de_Gennes} and Rossi and Pincus
\cite{Rossi} the correct scaling behavior was incorporated into the
free energy functional, whereas Semenov and coworkers have extended
the de Gennes model to determine finite chain length corrections
\cite{Semenov_96, Johner_96, Bonet_96, Semenov_96_review,
Semenov_98}. The latter calculations are of particular interest since
the polymer chain length is an important parameter in experiments
\cite{Flory_book} and computer simulations \cite{Binder, de_Joannis,
Jiminez}. Variation of the chain length thus provides a more stringent
testing of theoretical models \cite{Skau}.

A versatile model for the theoretical description of polymer
adsorption is the Scheutjens and Fleer
\cite{Scheutjens_Fleer,Scheutjens_Fleer_book} self-consistent field
model. The Scheutjens--Fleer model (SF) is a generalization of the
Flory--Huggins mean-field lattice model \cite{Flory_book} extended to
describe inhomogeneous polymer systems.  In the SF model, the Edwards
\cite{Edwards} self-consistent field equations for the polymer's Green
function are solved numerically on a lattice. The capability and
flexibility of the SF model has been demonstrated with its use to
calculate properties of inhomogeneous polymer systems containing
homopolymers, copolymers and polyelectrolytes and also surfactant
systems like micelles and membranes \cite{Fleer_book}. In recent
years, the SF algorithm has been made available in the multipurpose
computer program SFBOX \cite{van_Male_thesis}.

Recently, a self-consistent field formalism was derived for weakly
adsorbed polymers, valid for any chain length \cite{Blokhuis}.  One
notable difference from the classical ground state dominance result of
de Gennes \cite{de_Gennes_book} is the distal ordering of the
polymers; the finite chain length gives rise to oscillations on the
scale of the radius of gyration away from the surface. These
oscillations in the polymer concentration profile were first noted in
work by Semenov \emph{et al.} \cite{Semenov_96, Semenov_96_review},
and later observed \cite{Gucht} in the numerical solutions of the
Scheutjens and Fleer self-consistent field model. The onset of these
oscillations has also appeared in the context of analytical work and
computer simulations \cite{de_Joannis, Blokhuis, Bolhuis}.

In the present paper, we compare the free energy functional formalism
for weak adsorption \cite{Blokhuis} with the numerical results from
the Scheutjens and Fleer model. In particular we address the
observation made by van der Gucht \emph{et al.} \cite{Gucht} that in a
theta solvent, the oscillatory decaying function becomes a
monotonically decaying function when the bulk polymer density is below
a certain threshold value, similar to the Fisher--Widom transition
\cite{Fisher_Widom, Evans} in the damped oscillations found in a
simple liquid near a surface.  Our comparison will therefore consider
some general properties of the distal ordering in both good and theta
solvent conditions.

The outline of this article is as follows: In the next two sections we
discuss the self-consistent field theory for weak adsorption, in the
context of which our calculations are made, and the Scheutjens and
Fleer self-consistent field model. In Section 4, we make the
comparison between our analytical results and numerical calculations
carried out with the help of the computer program SFBOX. We end with a
discussion of results.

\section{Self-consistent field theory}

The Green function $G(\mathbf{r}, \mathbf{r}^{\, \prime}, N)$
describes the statistical weight of a single polymer chain of length
$N$ with one end at $\mathbf{r}$ and the other end at $\mathbf{r}^{\,
\prime}$.  The Green function is determined by the Edwards equation
\cite{Edwards}:
\begin{equation}
\frac{\partial}{\partial n} \, G(\mathbf{r}, \mathbf{r}^{\, \prime}, n) =
\frac{a^2}{6} \vec{\nabla}^2 \, G(\mathbf{r}, \mathbf{r}^{\, \prime}, n)
- \frac{U(\mathbf{r})}{k_{\rm B} T} \, G(\mathbf{r}, \mathbf{r}^{\, \prime}, n) \,,
\label{eq:Full_Edwards_equation}
\end{equation}
where $a$ is the polymer segment length and $U(\mathbf{r})$ is an as
yet unspecified external potential. As an initial condition to the
Edwards equation, we have that
\begin{equation}
\lim_{n \rightarrow 0} \, G(\mathbf{r}, \mathbf{r}^{\, \prime}, n) = 
\delta(\mathbf{r}-\mathbf{r}^{\, \prime}) \,.
\label{eq:Full_initial_condition}
\end{equation}
In terms of the Green function, one can construct the average segment
density:
\begin{equation}
\phi(\mathbf{r}) = a^3 \, N_{\rm p} \, \frac{
\int_0^N \!\! {\rm d}n \int \!\!{\rm d}\mathbf{r}^{\, \prime} \int \!\!
{\rm d}\mathbf{r}^{\, \prime\prime} \, G(\mathbf{r}^{\, \prime}, \mathbf{r}, n) \,
G(\mathbf{r}, \mathbf{r}^{\, \prime\prime}, N-n)}
{\int \!\! {\rm d}\mathbf{r}^{\, \prime} \int \!\! {\rm d}\mathbf{r}^{\, \prime\prime} \,
G(\mathbf{r}^{\, \prime}, \mathbf{r}^{\, \prime\prime}, N)} \,,
\label{eq:Full_segment_density}
\end{equation}
where $N_{\rm p}$ is the total number of polymer chains. This prefactor
determines the scale of the Green function; it is chosen such that the
density is equal to the given, uniform bulk density,
$\phi(\mathbf{r})\!=\!a^3 N_{\rm p} N/V \equiv \phi_b$, for a homogeneous
system.  Here, $\phi(\mathbf{r})$ is the polymer segment \emph{number}
density made dimensionless by the factor $a^3$ and can thus be
interpreted as the polymer segment \emph{volume} fraction.

The statistical weight $G(\mathbf{r},n)$ of a polymer chain with one
end at position $\mathbf{r}$ is obtained by integrating the Green
function over one of the chain's ends:
\begin{equation}
G(\mathbf{r}, n) \equiv \int \!\! {\rm d}\mathbf{r}^{\, \prime} \,
G(\mathbf{r}, \mathbf{r}^{\, \prime}, n) \,.
\label{eq:integrated_Green_function}
\end{equation}
For the case of polymer adsorption onto a planar, solid surface, the
segment density profile and Green function only depend on the
coordinate $z$ so that $\phi(\mathbf{r})\!=\!\phi(z)$ and
$G(\mathbf{r},n)\!=\!G(z,n)$. In the (isotropic) bulk region
$G(z,n)\!\rightarrow\!G_b(n)$, i.e.~independent of $z$. By solving the
Edwards equation (\ref{eq:Full_Edwards_equation}) with the initial
condition in Eq.(\ref{eq:Full_initial_condition}), one may show that
in the bulk region: $G_b(n)\!=\!e^{-U_b \, n/k_{\rm B} T}$,
where we have defined $U_b\!\equiv\!U(z\!=\!\infty)$. It is
convenient to redefine the statistical weight to absorb this trivial
$n$-dependence:
\begin{equation}
Z(z,n) \, \equiv \, e^{U_b \, n/k_{\rm B} T} \, G(z,n) \,.
\label{eq:definition_Z}
\end{equation}
The Edwards equation (Eq.(\ref{eq:Full_Edwards_equation})) and segment
density (Eq.(\ref{eq:Full_segment_density})) are then given by
\begin{eqnarray}
\frac{\partial}{\partial n} \, Z(z,n) &=&
\frac{a^2}{6} \frac{\partial^2}{\partial z^2} \, Z(z,n)
- \left[ \frac{U(z) - U_b}{k_{\rm B} T} \right] \, Z(z,n) \,,
\label{eq:Edwards_equation_z} \\
\phi(z) &=& \frac{\phi_b}{N} \, \int\limits_0^N \!\! {\rm d}n \,
Z(z,n) \, Z(z,N-n) \,,
\label{eq:segment_density_z}
\end{eqnarray}
with the initial condition:
\begin{equation}
\lim_{n \rightarrow 0} \, Z(z,n) = 1 \,.
\label{eq:initial_condition_z}
\end{equation}
In the self-consistent field model, the external potential $U(z)$ is
in turn expressed in terms of the polymer segment density with the
result that the set of equations
(\ref{eq:Edwards_equation_z})-(\ref{eq:segment_density_z}) becomes
self-consistently closed. Various forms for the self-consistent
external potential in terms of the segment density may be proposed.
It should in some way describe the interaction between different
polymer segments and of the polymer segments with the solid wall.  A
convenient form would be:
\begin{equation}
\frac{U(z)}{k_{\rm B} T} = v_0 \, \phi(z) + \frac{w_0^2}{2} \, \phi(z)^2
- m \, \phi^{\prime\prime}(z) + \frac{U_{\rm wall}}{k_{\rm B} T} \,,
\label{eq:U}
\end{equation}
The first two terms describe the polymer segment interaction in a
virial expansion where $v_0$, the so-called \emph{excluded volume}
parameter, is proportional to the second virial coefficient and
$w_0^2$ is proportional to the third virial coefficient of
segment--segment interactions. The third term describes the
\emph{non-local} character of the interaction and reflects the
non-homogeneous nature of the polymer segment distribution. Such a
term is analogous to the squared-gradient term in the van der Waals
theory for the liquid--vapor interface. The last term in
Eq.(\ref{eq:U}) gives the interaction with the wall. In the following
we will assume that it is short-ranged and that it can be approximated
by a delta function located \emph{at} the wall \cite{de_Gennes_book}:
\begin{equation}
\frac{U_{\rm wall}}{k_{\rm B} T} = - \frac{a^2}{6} \, \frac{1}{d} \, \delta(z) \,.
\label{eq:U_wall}
\end{equation}
The parameter $d\!>\!0$ is termed the extrapolation length
\cite{de_Gennes_book}; its inverse is a measure of the surface
interaction strength and leads to enhanced polymer adsorption,
$\phi(0)\!>\!\phi_b$. In the following, we will investigate the
situation where $1/d$ is small \cite{Blokhuis} leading to {\it weak
adsorption}.
\vskip 5pt
\noindent
{\bf Weak adsorption}
\vskip 5pt
\noindent
The assumption of weak polymer adsorption implies that
$\delta \phi(z)\!\equiv\!\phi(z)-\phi_b\!\ll\!\phi_b$ and
$\delta Z(z,n)\!\equiv\!Z(z,n) - 1\!\ll\!1$. Linearization of
Eqs.(\ref{eq:Edwards_equation_z}) and (\ref{eq:segment_density_z})
then leads to:
\begin{eqnarray}
\frac{\partial}{\partial n} \, \delta Z(z,n) &=&
\frac{a^2}{6} \frac{\partial^2}{\partial z^2} \, \delta Z(z, n)
- \frac{\delta U(z)}{k_{\rm B} T} \,,
\label{eq:Edwards_equation_dZ} \\
\delta \phi(z) &=& \frac{2 \phi_b}{N} \, \int\limits_0^N \!\! {\rm d}n \,
\delta Z(z,n) \,,
\label{eq:segment_density_dZ}
\end{eqnarray}
with the initial condition:
\begin{equation}
\lim_{n \rightarrow 0} \, \delta Z(z, n) = 0 \,.
\label{eq:initial_condition_dZ}
\end{equation}
The general form for the external potential expanded around
$\phi\!=\!\phi_b$ is given by
\begin{equation}
\frac{\delta U(z)}{k_{\rm B} T} = v \, \delta \phi(z)
- m \, \delta \phi^{\prime\prime}(z)
- \frac{a^2}{6} \, \frac{1}{d} \, \delta(z) \,.
\label{eq:U_linearized}
\end{equation}
The dimensionless parameter $v$ is a \emph{generalized excluded volume
parameter}. Since we have expanded around $\phi\!=\!\phi_b$ rather
than around $\phi\!=\!0$, as is done in the virial expansion in
Eq.(\ref{eq:U}), we have that $v\!=\!v(\phi_b)$ is not necessarily
equal to the second virial coefficient ($v\!\neq\!v_0$) and thus may
contain higher order interactions. For stability, we \emph{do} require
$v\!>\!0$; to describe the situation $v\!<\!0$, it would be necessary
to include more terms in the expansion around $\phi_b$. The above
expression for the external potential is used when we make the
comparison with the SF calculations.

For the situation that $m\!=\!0$, it was shown \cite{Blokhuis} that
the linearized set of equations in
Eqs.(\ref{eq:Edwards_equation_dZ})-(\ref{eq:U_linearized}) may be
reformulated into a free energy functional formalism with the free
energy functional given by
\begin{eqnarray}
\frac{a^3 \, F[\delta \phi]}{A \, k_{\rm B} T} &=& \frac{1}{8 \, \phi_b \, N}
\int\limits_{0}^{\infty} \!\! {\rm d}z \!\!
\int\limits_{0}^{\infty} \!\! {\rm d}z^{\prime} \,
\delta \phi^{\prime}(z) \, \delta \phi^{\prime}(z^{\prime})
\, \left[ \alpha_D(\frac{z-z^{\prime}}{2 R_{\rm g}})
- \alpha_D(\frac{z+z^{\prime}}{2 R_{\rm g}}) \right] \nonumber \\
&& + \frac{v}{2} \int\limits_{0}^{\infty} \!\! {\rm d}z
\left[ \, \delta \phi(z)^2 \, \right]
- \frac{a^2}{6} \, \frac{1}{d} \, \phi(0) \,.
\label{eq:free_energy_1}
\end{eqnarray}
where the polymer's radius of gyration $R_{\rm g}\!\equiv\!\sqrt{N a^2/6}$
and where
\begin{equation}
\alpha_D(x) \equiv \int\limits_{-\infty}^{\infty} \!\! \frac{{\rm d}k}{2
\pi} \, e^{i k x} \, \frac{k^2}{k^2 - 4 + 4 \, e^{-k^2/4}} \,.
\label{eq:definition_alpha_D}
\end{equation}
Minimization of the free energy in Eq.(\ref{eq:free_energy_1}) gives
for the segment density profile \cite{Blokhuis}
\begin{equation}
\delta \phi(z) = \frac{\phi_b \, \xi_b^2}{\pi \, d \, R_{\rm g}} \,
\int\limits_{0}^{\infty} \!\! {\rm d}t \, \cos(\frac{t z}{R_{\rm g}}) \,
\left[ \, \frac{t^2 - 1 + \exp(-t^2)}
{(\varepsilon/4) \, t^4 + t^2 - 1 + \exp(-t^2)} \, \right] \,,
\label{eq:solution_dphi}
\end{equation}
where we have introduced the parameter $\varepsilon$ as the (square
of) the bulk correlation length, $\xi_b\!\equiv\!a/\sqrt{3 \, v \,
\phi_b}$, divided by the polymer's radius of gyration:
\begin{equation}
\varepsilon \equiv \frac{\xi_b^2}{R_{\rm g}^2} = \frac{2}{v \, \phi_b \, N} \,.
\label{eq:varepsilon}
\end{equation}
The expression for the free energy can be generalized to the situation
$m\!\neq\!0$
\begin{eqnarray}
\frac{a^3 \, F[\delta \phi]}{A \, k_{\rm B} T} &=& \frac{1}{8 \, \phi_b \, N}
\int\limits_{0}^{\infty} \!\! {\rm d}z \!\!
\int\limits_{0}^{\infty} \!\! {\rm d}z^{\prime} \,
\delta \phi^{\prime}(z) \, \delta \phi^{\prime}(z^{\prime})
\, \left[ \alpha_D(\frac{z-z^{\prime}}{2 R_{\rm g}})
- \alpha_D(\frac{z+z^{\prime}}{2 R_{\rm g}}) \right] \nonumber \\
&& + \frac{1}{2} \int\limits_{0}^{\infty} \!\! {\rm d}z
\left[ \, v \, \delta \phi(z)^2 + m \, \delta \phi^{\prime}(z)^2 \right]
- \frac{a^2}{6} \, \frac{1}{d} \, \phi(0) \,.
\label{eq:free_energy_2}
\end{eqnarray}
The resulting segment density profile is then given by
\begin{equation}
\delta \phi(z) = \frac{\phi_b \, \xi_b^2}{\pi \, d \, R_{\rm g}} \,
\int\limits_{0}^{\infty} \!\! {\rm d}t \, \cos(\frac{t z}{R_{\rm g}}) \,
\left[ \, \frac{t^2 - 1 + \exp(-t^2)}
{(\varepsilon/4) \, t^4 + (1 + \beta \, t^2) \, (t^2 - 1 + \exp(-t^2))}
\, \right] \,,
\label{eq:solution_dphi_m}
\end{equation}
where we have introduced the parameter $\beta$ as the (square of) the
length defined by the ratio of $m$ and $v$,
$\xi_{\rm m}\!\equiv\!\sqrt{m/v}$, divided by the polymer's radius of
gyration:
\begin{equation}
\beta \equiv \frac{\xi_{\rm m}^2}{R_{\rm g}^2} = \frac{6 \, m}{N a^2 \, v} \,.
\label{eq:beta}
\end{equation}
The length $\xi_{\rm m}$ is the typical length scale connected to spatial
inhomogeneities in the segment density caused by the interactions
between segments excluding interactions due to the chain's
connectivity (which are described by $\xi_b$). In a good solvent
$\xi_b$ dominates over $\xi_{\rm m}$ while $\xi_{\rm m}$ gains in significance
near the theta region.

The main goal in this article is to compare the segment density
profiles given in Eqs.(\ref{eq:solution_dphi}) and
(\ref{eq:solution_dphi_m}) to numerical results from the Scheutjens
and Fleer lattice model.

\section{Scheutjens and Fleer self-consistent field model}

The Scheutjens--Fleer model
\cite{Scheutjens_Fleer,Scheutjens_Fleer_book} gives an efficient way
to solve, self-consistently, the discrete version of the Edwards
equation (\ref{eq:Edwards_equation_z}) on a lattice. The lattice
distance is taken to be equal to the polymer segment length $a$, and
$1/\lambda$ is the lattice coordination number that we set equal to
$1/\lambda\!=\!6$ corresponding to a cubic lattice. The distance $z$
away from the surface is here a discrete variable, $z\!=\!0, 1,
\ldots, M$ layers (See Figure 1). A hard surface is located at
$z\!=\!0$, and at $z\!=\!M$ we have reflective boundary conditions
($M$ should be chosen large enough for the system to reach bulk
density).

The polymer is modeled as a one-dimensional walk perpendicular to the
surface, where the walk can either move to the neighboring layer or
stay in the layer it already is. Each step is weighted with the
probability of going to a neighboring layer ($\lambda$) or staying in
the same layer ($1 - 2 \lambda$). This gives the following recurrence
relation for the Green function $G(z,s)$ (the discrete analog of
$Z(z,n)$), which is the statistical weight of finding the end segment
$s$ of a chain in layer $z$,
\begin{eqnarray}
G(z,s+1) &=& G(z) \, \left[ \, \lambda \, G(z-1,s) + (1 - 2 \lambda) \, G(z,s)
+ \lambda \, G(z+1,s) \, \right] \,, \nonumber \\
         &\equiv& G(z) \, <\! G(z,s) \!> \,.
\label{eq:recurrence_relation}
\end{eqnarray}
As a starting point for the recurrence relation we have that
$G(z,1)\!=\!G(z)$, the statistical weight of a single polymer
segment. The brackets $<\! \ldots \!>$ denote the neighbor-weighted
average as defined above. In the first layer, $z\!=\!1$, the
neighbor-weighted average is defined as:
\begin{equation}
<\! G(1,s) \!> \, \equiv \, (1 - 2 \lambda) \, G(1,s) + \lambda \,
G(2,s) \,.
\label{eq:<1>}
\end{equation}

In analogy with the continuous case, the segment density profile is
obtained by summing over all possible paths of a polymer of length $N$
passing through position $z$:
\begin{equation}
\phi(z) = \frac{\phi_b}{N}
\sum_{s=1}^N \frac{G(z,s) \, G(z,N-s+1)}{G(z)} \,, 
\end{equation}
where the factor $G(z)$ in the denominator accounts for the double
counting of the statistical weight at layer $z$. The statistical
weight $G(z)$ is written as the Boltzmann factor of the segment
potential
\begin{equation}
G(z) = e^{-U_{\rm SF}(z)/k_{\rm B} T} \,.
\label{eq:G(z)}
\end{equation}
The segment potential $U_{\rm SF}(z)$ accounts for the interactions
between polymer segments and the surface-monomer interaction. The zero
of the potential is at $z\!=\!\infty$ so that $U_{\rm
SF}(z\!=\!\infty)\!=\!0$ and $G(z\!=\!\infty)\!=\!1$. In the
Scheutjens and Fleer model the form for the segment potential is
chosen such that it is consistent with the Flory--Huggins free energy
expression for a homogeneous polymer solution \cite{Flory_book}.  It
is given by \cite{Scheutjens_Fleer}:
\begin{equation}
\frac{U_{\rm SF}(z)}{k_{\rm B} T} = - \ln \left( \frac{1 - \phi(z)}{1 - \phi_b} \right)
- \chi \, \left( <\! \phi(z) \!> - <\! 1 - \phi(z) \!> + 1 - 2 \phi_b \right) 
- \chi_{\rm s}\, \delta_{z,1} \,.
\label{eq:U_SF}
\end{equation}
The first term is derived from the translational entropy of the
``solvent molecules'' (the sites not occupied by the polymer segments)
in the Flory--Huggins theory. The second term describes the effective
monomer--monomer interaction energy through the Flory parameter $\chi$
\cite{Flory_book}. (Notice that this term is \emph{non-local} since
the interactions between monomers in different layers is explicitly
taken into account by the use of $<\!\phi(z)\!>$ instead of
$\phi(z)$.)  The final term describes the surface-monomer interaction
through the Silberberg parameter $\chi_{\rm s}$
\cite{Silberberg}. This parameter corresponds to the energy gain for a
polymer segment to replace a solvent molecule at the surface
($z\!=\!1$).

As input parameters for the SF calculations we thus have: $\phi_b$,
$N$, $\chi$, and $\chi_{\rm s}$. Many extensions to this model are
possible and many have been made \cite{Fleer_book,
van_Male_thesis}. One may, for example, introduce different polymer
architectures, inhomogeneities in more than one dimension, and curved
lattice geometries, but the underlying principles remain the same.

\section{Comparison of results}

To make a comparison between the density profiles from the analytical
theory and the Scheutjens--Fleer self-consistent field model, one
needs to relate the parameters in the analytical theory to those in
the lattice model. In particular, one needs to: (\textit{i}) relate
the interaction parameters $v$ and $m$ to the Flory--Huggins parameter
$\chi$, and (\textit{ii}) relate the extrapolation length $d$ (or
rather $1/d$) to the surface interaction parameter $\chi_{\rm s}$.

The discrete self-consistent field equations in the SF model can be
transformed into continuous equations by assuming that the variations
with $z$ are small. In Appendix A we show that the recurrence relation
in Eq.(\ref{eq:recurrence_relation}) then reduces to the continuous
Edwards equation (\ref{eq:Edwards_equation_z}) with the external
potential equal to $U_{\rm SF}(z)$. To make the correspondence between
$\chi$ and the parameters $v$ and $m$, one therefore needs to expand
the expression for $U_{\rm SF}(z)$ in Eq.(\ref{eq:U_SF}) around
$\phi\!=\!\phi_b$, using that $<\!\phi\!>\rightarrow\!\phi + (a^2/6)
\, \phi^{\prime\prime}$ in the continuous limit (cf. Appendix A), and
compare the result with $\delta U(z)$ in
Eq.(\ref{eq:U_linearized}). One then finds that
\begin{equation}
v \, \longleftrightarrow \, \frac{1}{1 - \phi_b} - 2 \, \chi \,, \hspace*{40pt}
m \, \longleftrightarrow \, \frac{a^2}{3} \, \chi \,.
\label{eq:v_and_m}
\end{equation}
%

It is less straightforward to arrive at a relation between the
extrapolation length $d$ and the surface interaction parameter
$\chi_{\rm s}$. It should be realized that the presence of a solid
surface is treated fundamentally different in the two models.  In the
weak adsorption model, the treatment of the presence of the solid
surface is derived from the de Gennes model for polymer adsorption
\cite{de_Gennes_book}. In the de Gennes model, the free energy
functional is strictly defined for $z\!>\!0$; an interaction energy is
added located at $z\!=\!0$ (Eq.(\ref{eq:U_wall})), but such a term
merely results in a boundary condition to the differential equation
obtained from minimizing the free energy functional. As a result, the
polymer segment density does not necessarily go to zero \emph{at} the
wall, $\phi(0)\!\neq\!0$.

In the SF model, it is implicitly assumed that the polymer segment
density at the wall is zero. This can be read off from the definition
of the neighbor-weighted average in the first layer (see
Eq.(\ref{eq:<1>})), which for the segment density reads:
\begin{equation}
<\! \phi(1) \!> \, = \, (1 - 2 \lambda) \, \phi(1) + \lambda \, \phi(2) \,,
\end{equation}
i.e.~ $\phi(0)\!=\!0$. It turns out that a consequence of this
condition is that \emph{enhanced} adsorption only occurs in the SF
model above some threshold value, $\chi_{\rm s}\!>\!\chi_{\rm sc}$.
The situation $\chi_{\rm s}\!=\!\chi_{\rm sc}$ then corresponds to the
condition $1/d\!=\!0$, which gives $\phi(z)\!=\!\phi_b$ for all $z$ in
the de Gennes model. The threshold value as a function of $\chi$ may
therefore be determined by using that at the threshold one should have
$\phi(z)\!\approx\!\phi_b$ for all $z$. For very long chains one
finds: $\chi_{\rm sc}\!=\!-\lambda \, \chi - \ln(1 - \lambda)$
\cite{Fleer_van_Male}.  Next, by comparing the expressions for the
wall interaction terms in the two models, Eqs.(\ref{eq:U_linearized})
and (\ref{eq:U_SF}), one arrives at the following identification
\cite{Johner_96, Fleer_van_Male}
\begin{equation}
\frac{a^2}{6} \, \frac{1}{d} \, \delta(z) \, \longleftrightarrow \, 
\left( \chi_{\rm s} - \chi_{\rm sc} \right) \, \delta_{z,1} \hspace*{20pt}
\Longrightarrow \hspace*{20pt}
\frac{a}{d} = 5 \, \left[ \, \chi_{\rm s} + \frac{\chi}{6} - \ln(6/5) \, \right] \,,
\label{eq:relation_d_to_chi_s}
\end{equation}
where we used that $\delta_{z,1}\!\longleftrightarrow\!a \, (1-\lambda) \, \delta(z)$
\cite{Fleer_van_Male}.

We now have the necessary ingredients to compare the analytic
expressions for the segment density profiles in
Eqs.(\ref{eq:solution_dphi}) and (\ref{eq:solution_dphi_m}) to the
numerical results of the SF model.  It is good to realize that the
inverse of the extrapolation length, $1/d$, only shows up as a
prefactor to the analytic expressions for the segment density
profile. Therefore, if one is interested in only the
\emph{qualitative} features of the segment density profile, knowledge
of the precise value of $1/d$ is not needed. Only when one wants to
make a quantitative comparison with the SF model, it is necessary to
use the above relation between $1/d$ and $\chi_{\rm s}$.

Next, we first consider the case of an \emph{athermal} solvent
($\chi\!=\!0$) and compare the SF model results to
Eq.(\ref{eq:solution_dphi}). In particular, we will investigate the
distal oscillations for weak and strong polymer adsorption. Second, we
consider a theta solvent ($\chi\!=\!1/2$), as was previously done by
van der Gucht \emph{et al.} \cite{Gucht}, who showed that when the
bulk density is below some threshold value, the oscillations in the
distal profile disappear similar to the Fisher--Widom transition
\cite{Fisher_Widom, Evans} in the damped oscillations for a simple
liquid near a surface.  A quantitative comparison is made with the
analytical expression for the segment density profile in
Eq.(\ref{eq:solution_dphi_m}). Finally, we consider the location of
the Fisher--Widom transition with respect to the complete
Flory--Huggins phase diagram \cite{Flory_book} and also make a
comparison with MC simulations \cite{de_Joannis}.

\subsection{Athermal solvent}

For an \emph{athermal} solvent we have that $\chi\!=\!0$ so that:
\begin{equation}
v = \frac{1}{1 - \phi_b} \approx 1 \,, \hspace*{40pt}
m = 0 \,.
\end{equation}
In Figure 2a, the polymer segment density profile as calculated by the
SF model is shown for two polymer chain lengths ($N\!=\!1000$ and
$N\!=\!10000$) in the case of weak polymer adsorption ($\chi_{\rm
s}\!=\!0.20$). To show $\delta \phi(z)/\phi_b$ on a logarithmic scale,
here, and in later plots, the absolute value has been taken.  Also
shown as the solid line is the ground state dominance segment density
profile \cite{de_Gennes_book}:
\begin{equation}
\phi(z) = \frac{\phi_b}{\tanh^2(z/\xi_b + x_0)} \,,
\label{eq:cotanh-profile}
\end{equation}
with the integration constant $x_0\!=\!(1/2) \, {\rm arcsinh}(2 d/\xi_b)$. 
In the comparison of Figure 2a, the expression for $a/d$ in
Eq.(\ref{eq:relation_d_to_chi_s}) was used.

The profiles corresponding to the two chain lengths are close to each
other and close to the ground state dominance profile in this distance
range. At very large distances, however, markedly different behavior
can be observed; in Figure 2b, the same SF model results as in Figure
2a are plotted, showing oscillations in the segment density profile on
the order of the polymer's radius of gyration. The two solid lines in
Figure 2b are the density profiles calculated by
Eq.(\ref{eq:solution_dphi}), using Eq.(\ref{eq:relation_d_to_chi_s}).
Without any adjustable parameters, a strikingly good quantitative
agreement between the SF model and Eq.(\ref{eq:solution_dphi}) is
obtained.

Even for strong adsorption the density profile in
Eq.(\ref{eq:solution_dphi}) is expected to give an accurate
description for distances far away from the surface, since also there
we have that $\delta \phi(z)\!\ll\!\phi_b$. However, since the density
profile differs at short distances, the boundary condition may need to
be replaced by some \emph{effective} boundary condition. The result is
that although the \emph{shape} of the density profile is correctly
described by Eq.(\ref{eq:solution_dphi}) for strong adsorption, it may
need to be shifted in the $z$-direction introducing the shift as an
additional fit parameter. In Figure 3, the polymer segment density
profile as calculated by the SF model is shown for a long polymer
chain ($N\!=\!1000$) and for two values of the surface interaction
parameter, $\chi_{\rm s}\!=\!0.25$ and $\chi_{\rm s}\!=\!1$,
corresponding to strong adsorption. The solid lines are the density
profiles calculated by Eq.(\ref{eq:solution_dphi}), using
Eq.(\ref{eq:relation_d_to_chi_s}) to set the scale of the $y$-axis.
Furthermore, the density profiles are shifted in the $z$-direction by
$-1.5 \, a$ (lower solid line) and $-4.5 \, a$ (top solid line). The
asymptotic profiles are seen to indeed match quite well.

\subsection{Theta solvent}

We next consider the case of a $\Theta$-solvent ($\chi\!=\!1/2$) where
we have that 
\begin{equation}
v = \frac{1}{1 - \phi_b} - 1 \approx \phi_b \,,
\hspace*{40pt} m = \frac{a^2}{6} \,.
\end{equation}
It is good to realize that theta solvent conditions do not correspond
to a vanishing generalized excluded volume parameter ($v\!\neq\!0$).

In Ref. \cite{Gucht} the observation was made that the oscillations,
prominently present in Figures 2 and 3, disappear in a
$\Theta$-solvent when the density is below a certain threshold bulk
density, similar to the Fisher--Widom transition \cite{Fisher_Widom}
for a simple liquid. In Figure 4, we show the characteristic signature
of the Fisher--Widom (FW) transition going from $\phi_b\!=\!0.01$
(solid line) to $\phi_b\!=\!0.001$ (dashed line), for $N=1000$ and for
the same surface interaction strength, $\chi_{\rm s}\!=\!-1/12$, as in
Ref. \cite{Gucht}. (Note that this value is below the adsorption
threshold value, so that we are dealing with polymer depletion instead
of adsorption.)  To understand this qualitative change in the behavior
of the segment density profile, we further investigate the expression
for the density profile in Eq.(\ref{eq:solution_dphi_m}).

In Appendix B, it is shown that the asymptotic behavior is determined
by the poles in the complex plane of the integrand's denominator of
the expression for the density profile in
Eq.(\ref{eq:solution_dphi_m}). The resulting expression for the
asymptotic segment density profile as given by
Eq.(\ref{eq:asymptotic_behaviour_dphi_sum}) shows that the relevant
poles correspond to either a purely exponentially decaying function:
\begin{equation}
\delta \phi(z) \propto \exp(- \frac{A_0 z}{R_{\rm g}}) \,,
\label{eq:asymptotic_behaviour_dphi_exp}
\end{equation}
or to an exponentially decaying sinusoid:
\begin{equation}
\delta \phi(z) \propto \exp(- \frac{A_1 z}{R_{\rm g}}) \,\,
\sin(\frac{B_1 z}{R_{\rm g}} + \varphi) \,.
\label{eq:asymptotic_behaviour_dphi_osc}
\end{equation}
Which of these two functions dominates the asymptotic behavior depends
on the relative magnitudes of $A_0$ and $A_1$; when $A_0\!>\!A_1$, the
segment density decays asymptotically as an exponentially decaying
sinusoid whereas it decays exponentially when $A_0\!<\!A_1$.  The
value of the coefficients $A_0$, $A_1$ and $B_1$, depend on the value
of the parameters $\varepsilon$ and $\beta$. In Figure 5a, we have set
$\beta\!=$ 0.18 and plotted $A_0$, $A_1$ and $B_1$ as a function of
$\varepsilon$.  At a certain value of $\varepsilon$ (in this example:
$\varepsilon^{\rm FW}\!\approx\!$ 6.464), $A_0$ and $A_1$ cross
marking the location of the Fisher--Widom transition (arrow).  By
varying $\beta$, the whole locus of Fisher--Widom transitions can then
be traced in this way (see Figure 5b).

To compare the analytically determined Fisher--Widom transition in
Figure 5b with the results from the SF model, we located the
Fisher--Widom transition by determining the bulk density where the
\emph{first} oscillation in the numerical density profiles disappears.
This is the point where the first minimum (located closest to the
wall) of the oscillating density profile crosses the value of the bulk
density. This procedure gives only an approximation to the location of
the real Fisher--Widom transition. A closer investigation reveals that
the density oscillations do not disappear all at once at the same
polymer bulk density; the first oscillation disappears before the
second oscillation with decreasing bulk density, and presumably before
all oscillations beyond that. Still, since the bulk densities at which
subsequent oscillations vanish differ only slightly, we believe this
procedure to give a fairly accurate approximation.

In Figure 6, the polymer segment density profile is shown for a
$\Theta$-solvent with $N\!=\!10000$ for various bulk densities just
below and above the Fisher--Widom transition: $\phi_b\!=$ 0.00078
(solid line), $\phi_b\!=$ 0.00082 (dashed line), $\phi_b\!=$ 0.00086
(dot-dashed line), and $\phi_b\!=$ 0.00090 (dotted line). In (a), the
density profiles are calculated in the SF model; in (b) the density
profile is calculated using
Eq.(\ref{eq:asymptotic_behaviour_dphi_sum}). Good quantitative
agreement is obtained.

In Figure 7, we have located the Fisher--Widom transition in the SF
model by determining the bulk density where the first oscillation
disappears for different chain lengths (circles). The surface
interaction strength $\chi_{\rm s}\!=\!-1/12$, as in
Ref. \cite{Gucht}. The numerical data are consistent with a scaling of
the bulk density $\phi_b^* \sim 1/N$. The solid line is the
Fisher--Widom transition shown in Figure 5b, where we have used that
$N\!=\!\varepsilon/(2\beta^2)$ and $\phi_b\!=\!2 \beta/\varepsilon$ to
transform from ($\varepsilon$, $\beta$) to ($N$, $\phi_b$) as
variables. Good agreement is obtained with the SF calculations.

Also shown in Figure 7, as the squares, are the SF calculations
results for $N\!=\!200$ and $N\!=\!1000$ reported in
Ref. \cite{Gucht}, which differ somewhat from our results. Since these
SF calculations are done with exactly the same set of parameters used
to determine our results represented by circles, this difference must
be attributed to the different procedure used in Ref. \cite{Gucht} to
locate the Fisher--Widom transition. In the analysis in
Ref. \cite{Gucht} the segment density profile is taken to be of the
following form:
\begin{equation}
\frac{\delta \phi(z)}{\phi_b} = \tanh^2(\frac{z-z_0}{\xi_{b, \rm num}})
- 1 + C_{\rm num} \, \exp(- \frac{A_{\rm num} \, z}{R_{\rm g}}) \,\,
\sin(\frac{B_{\rm num} \, z}{R_{\rm g}} + \varphi_{\rm num}) \,.
\label{eq:dphi_Gucht}
\end{equation}
Then, by locating, as a function of $\phi_b$, the position where
$\xi_{b, \rm num}/2$ equals $R_{\rm g}/A_{\rm num}$, the Fisher--Widom
transition is determined.

Even though the analysis in Ref. \cite{Gucht} is done in a more
judicious manner than simply fitting the numerical segment density
profile to Eq.(\ref{eq:dphi_Gucht}) (see Ref. \cite{Gucht} for
details), the large number of parameters involved makes it difficult
to accurately determine the location of the Fisher--Widom
transition---especially since at the transition the relevant length
scales cross. For instance, in the example shown in Figure 4, the
Fisher--Widom transition as determined by the bulk density where the
first oscillation disappears, is located at $\phi^{\rm
FW}_b\!\approx\!0.00707$, whereas the value reported in
Ref. \cite{Gucht} is $\phi^{\rm FW}_b\!\approx\!0.032$. This result
would imply that the density profile depicted by the solid line in
Figure 4 for $\phi_b\!=\!0.01$ lies in the purely exponentially
decaying-region, which seems unlikely.

Nevertheless, it is expected that this fitting procedure is more
reliable in the region where the different length scales are
well-separated, e.g. when $\phi_b\!\gg\!\phi^{\rm FW}_b$. In this
region, the fit parameters were numerically determined as \cite{Gucht}
\begin{equation}
A_{\rm num} \simeq \frac{1}{0.19 \,\, \sqrt{6}} \simeq 2.1 \,,
\hspace*{20pt}
B_{\rm num} \simeq \frac{2 \pi}{1.5 \,\, \sqrt{6}} \simeq 1.7 \,,
\hspace*{20pt}
\xi_{b, \rm num} \simeq \frac{0.52 \, a}{\phi_b} \,.
\label{eq:A_B_ksi_numerics}
\end{equation}
These values can be compared to the analytical density profile in the
corresponding regime: $\varepsilon\!\ll\!1$ and $\beta\!\ll\!1$. For
$\beta\!=\!0$ and expanding in $\varepsilon$, it was shown that the
segment density profile is then given by \cite{Blokhuis}
\begin{equation}
\frac{\delta \phi(z)}{\phi_b}\ = \frac{\xi_b}{d} \, e^{- 2 z/\xi_b} +
C \, \exp(- \frac{A \, z}{R_{\rm g}}) \,\, \sin(\frac{B \, z}{R_{\rm g}} +
\varphi) \,,
\label{eq:dphi_small_beta_and_epsilon}
\end{equation}
with
\begin{equation}
A = 2.217792 \ldots \,, \hspace*{25pt} B = 1.682188 \ldots \,,
\hspace*{25pt} \xi_b = \frac{1}{\sqrt{3} \, \phi_b} \simeq \frac{0.58
\, a}{\phi_b} \,.
\label{eq:A_B_ksi_analytical}
\end{equation}
Thus, we find in the region $\phi_b\!\gg\!\phi^{\rm FW}_b$ good
agreement between the analytical results \cite{Blokhuis} and the
fitting parameters in Eq.(\ref{eq:A_B_ksi_numerics}).

\subsection{Phase diagram}

Since the Fisher--Widom transition found in a $\Theta$-solvent seems
to be absent in an athermal solvent, it is worthwhile to locate the
Fisher--Widom transition as a function of solvent quality.

In Figure 8, the Fisher--Widom transition is shown with respect to the
full polymer phase diagram for $N\!=\!100$. As a reference, the
Flory--Huggins spinodal and binodal regions (dashed lines) are also
shown \cite{Flory_book}. The symbols are the SF results for the
Fisher--Widom transition for enhanced polymer adsorption $\chi_{\rm
s}\!=\!0.5$ (circles) and for polymer depletion $\chi_{\rm s}\!=\!0$
(squares). The solid line is the Fisher--Widom transition plotted in
Figure 5b transformed from ($\varepsilon$, $\beta$) to ($\chi$,
$\phi_b$) as variables, using Eq.(\ref{eq:v_and_m}). Good agreement is
obtained with the SF calculations, independent of the polymer-wall
interaction strength.

In the derivation of the analytical result for the Fisher--Widom
transition, we have assumed that the generalized excluded volume
parameter is always positive, $v\!>\!0$. To verify that the
Fisher--Widom transition lies within the region of applicability, the
locus of points where $v\!=\!0$ (which gives $2
\chi\!=\!1/(1-\phi_b)$) is shown in Figure 8 as the dotted
line. Furthermore, the approximate location of the dilute to
semi-dilute transition ($\phi^* \!=\! 1/(v N)$ \cite{Daoud_Jannink},
with $v$ given by Eq.(\ref{eq:v_and_m})) is shown as the dot-dashed
line in Figure 8. Since both the mean-field model presented here as
well as the SF model are expected to be reliable only in semi-dilute
and concentrated solutions, the location of the Fisher--Widom
transition shown is physically most relevant in the theta region.

As a final remark, we mention that a useful approximation to the
location of the Fisher--Widom transition can be obtained by
approximating $\beta^{\rm FW}$ by its $\varepsilon\!\rightarrow\!0$
value: $\beta^{\rm FW}\!\approx\!0.20$ (see Figure 5b). This gives $2
\chi^{\rm FW}\!\approx\!N/[(N+5) \, (1 - \phi_b)]$. For very long
polymer chains, the Fisher--Widom transition thus approaches the line
where $v\!=\!0$.

\subsection{Comparison with Monte Carlo simulations}

Recently, de Joannis \emph{et al.} \cite{de_Joannis} performed Monte
Carlo (MC) simulations of polymer chains at surfaces in an athermal
solvent. In Figure 9a, the reduced segment density
$\delta\phi(z)/\phi_b$ is shown as a function of $z/R_{\rm g}$ (Figure 9b
shows the same profiles on a logarithmic scale).  For the MC
simulations (circles), the following parameter values were used: chain
length $N\!=\!200$, bulk density $\phi_b\!=\!0.0216$, radius of
gyration $R_{\rm g}/a\!=\!9.76$, and surface interaction strength
$\varepsilon_s\!=$ 1.0 $k_{\rm B} T$ \cite{de_Joannis}. The squares
are the results from the SF model for the same parameters
($N\!=\!200$, $\chi\!=\!0$, bulk density $\phi_b\!=\!0.0216$) taking
$\chi_{\rm s}\!=\!1.0$ (variation of $\chi_{\rm s}$ does not lead to
very different results).  Shown as the solid and dashed line is the
density profile calculated using Eq.(\ref{eq:solution_dphi}) for
$a/d\!=\!7.35$ and $a/d\!=\!4.09$, respectively (both profiles are
shifted in the $z$-direction to fit the data).  The value of
$a/d\!=\!7.35$ was chosen such that the depth of the minimum in the
segment density profile matches that of the MC simulations, whereas
the value of $a/d\!=\!4.09$ was chosen to correspond to $\chi_{\rm
s}\!=\!1.0$ (using Eq.(\ref{eq:relation_d_to_chi_s})). The SF model
corresponds well with the MC results although the depth of the minimum
differs by a factor of two. The asymptotic formula in
Eq.(\ref{eq:solution_dphi}) is able to describe the MC-data (two
fitting parameters) and SF model (one fitting parameter) well. So,
also in the Monte Carlo simulations oscillations in the polymer
concentration profile can be observed, even though it is not possible
from the numerical data (and it is hard to get more accurate profiles)
to observe more than the first oscillation.

\section{Discussion}

In this article, we have compared the analytic expressions for the
segment density profiles obtained in the context of the weak
adsorption model \cite{Blokhuis} to the numerical results of the
Scheutjens--Fleer model \cite{Scheutjens_Fleer,Scheutjens_Fleer_book}.
The analytic expressions are expected to be valid when the polymer
adsorbs only weakly to the solid surface or, for strong adsorption,
when the distance to the surface is large. Indeed, excellent
quantitative agreement for the distal, oscillatory density profile was
obtained for very good (athermal) solvent conditions, for both weak
and strong polymer adsorption (Figures 2 and 3).

We have further investigated the previously observed Fisher--Widom
transition in SF model calculations by van der Gucht \emph{et al.}
\cite{Gucht} for a $\Theta$-solvent. We demonstrated that the location
of the Fisher--Widom transition can be described by the weak
adsorption model when the non-local character of the monomer--solvent
interaction is taken into account. The quantitative agreement that is
then achieved (see Figures 6, 7 and 8), gives confidence in
identifying the non-local character of the interactions in the SF
model as the source of the Fisher--Widom transition observed
\cite{Gucht}. To further substantiate this identification, we also
carried out separate SF calculations in which the non-local
interactions were removed by replacing $<\!\phi(z)\!>$ by $\phi(z)$ in
Eq.(\ref{eq:U_SF}). In this way, we could explicitly verify
\cite{Skau_thesis} that without the non-local interactions present,
the oscillations always remain with decreasing bulk polymer
concentration while maintaining the quantitative agreement with the
segment density profiles of the weak adsorption model.

Although it is shown that the introduction of a non-local interaction
term is useful for the comparison with the SF model, one should
address the relevance of it when comparing with real, experimental
polymer systems. Both the model presented here as well as the SF model
are based on the mean-field assumption which is expected to be
reasonable for semi-dilute and concentrated solutions (overlapping
polymer coils) but is expected to fail for dilute solutions (isolated
coils), where fluctuations and inhomogeneities are more
pronounced. Under good solvent conditions the Fisher--Widom transition
occurs at polymer concentrations well below the dilute to semi-dilute
transition, $\phi^{\rm FW}_b \ll \phi^* \sim 1/(v_0 N)$
\cite{Daoud_Jannink}, and the mean-field results for the location of
the Fisher--Widom transition are probably not so reliable. In theta
and poor solvent conditions, close to the spinodal decomposition
region (theta region), the Fisher--Widom transition shifts to much
larger concentrations and occurs at a concentration comparable to the
dilute to semi-dilute transition concentration, $\phi^{\rm FW}_b
\simeq \phi^* \sim 1/(w_0 \, N^{1/2})$ \cite{Daoud_Jannink} (see
Figure 8). We therefore expect our mean-field results to be most
relevant to describe the occurrence and location of the Fisher--Widom
transition in experimental polymer systems in the theta region.

It may seem somewhat surprising that the non-local interactions in the
SF model, whose range is only one lattice distance $a$, may interfere
with the oscillatory behavior on the scale of the polymer's radius of
gyration. However, the length scale connected with the non-local
character is rather $\xi_{\rm m}\!\equiv\!\sqrt{m/v}$ which diverges
(similar to the bulk correlation length near the critical point in an
ordinary liquid) when $v$ is small, i.e.~when the solvent is poor
enough.

The Fisher--Widom transition in a simple liquid occurs when the
attractive part of the interaction potential between the molecules
dominates over the repulsive part \cite{Fisher_Widom, Evans}. For high
densities the hard core interaction gives damped oscillations away
from the surface, while these oscillations disappear when the
inter-molecular attraction balances the repulsion. The similarity with
the transition observed for polymers is striking; the $\Theta$-point
defines the system conditions where the monomer attraction cancels the
monomer repulsion. For a good solvent the effective monomer
interaction is repulsive and oscillations are predicted for all
densities \cite{Blokhuis}. There are, however, nontrivial differences
between monomers and liquid molecules; the most important difference,
of course, is the polymer chain connectivity. The chain connectivity
gives rise to oscillations with a wavelength of the size of polymer
coil, while in a simple liquid the wavelength is given by the
molecular radius. It is tempting to think of the polymer solution as a
``liquid'' of polymer coils close to the surface. But this intuitive
idea fails to consider the strong interdigitation of the coils at
semi-dilute concentrations and that the coil changes its conformation
and size close to the surface (while a liquid molecule keeps its hard
core shape). One should therefore be careful to make too much out of
such analogies.

In the comparison of the Scheutjens and Fleer model
\cite{Scheutjens_Fleer,Scheutjens_Fleer_book} to analytical theories
such as the de Gennes model \cite{de_Gennes_book, de_Gennes}, and
extensions thereof such as the weak adsorption model \cite{Blokhuis},
a number of differences can be distinguished. A clear difference is
the fact that the SF model is a lattice model with the obvious
consequence that the parameter describing the distance to the
substrate is discrete; $z\!=\!1, 2, 3, \ldots$. The discreteness of
space is most consequential in the very vicinity of the wall, where
the monomer length scale becomes important.

Another important difference between the SF model and the de
Gennes-like models concerns the boundary condition at the wall. In
these latter models, the free energy functional is only strictly
defined for $z\!>\!0$. An interaction energy is defined located at
$z\!=\!0$, but the addition of such a term merely serves as a boundary
condition to the differential equation which results from the
minimization of the free energy functional. The result is that the
polymer segment density not necessarily goes to zero \emph{at} the
wall, $\phi(0)\!\neq\!0$.  Even though in real polymer systems, the
polymer segment density should become zero (or exponentially close to
it) at some point, it is expected that a microscopic shift in the
precise location of the wall---this location is not well-defined
within a microscopic distance anyway---should restore the validity of
the de Gennes-like models at least for enhanced polymer adsorption
\cite{de_Gennes}.

In the SF model, the free energy functional is already defined for
whole space from the start. The reason that the density profile is
zero when $z\!\leq\!0$ is due to the explicit addition of an external
potential which gives an infinite energy penalty for $z\!\leq\!0$.
Apart from this external potential there may be an energy contribution
to the first layer at $z\!=\!1$ which will generally be attractive.
The result of the infinite repulsion at $z\!\leq\!0$ is that the
polymer density is zero at the wall $\phi(0)\!=\!0$. The presence of
this extra boundary condition makes the SF model essentially different
from the de Gennes-like models, but similar to the single chain
mean-field theory by Szleifer and others \cite{Szleifer, Bonet} and
computer simulations \cite{Binder, de_Joannis, Jiminez}. This boundary
condition (but without the presence of any additional attraction) is
also used in the calculations by Eisenriegler and coworkers
\cite{Eisenriegler_book, Eisenriegler_Dietrich, Maasen}. Still, it
seems, as we have done here, that a meaningful comparison between the
SF model and the de Gennes-like models can still be made when the
adsorption energy in the first layer is above some given threshold
value.

\appendix
\section{Discrete and continuous Edwards equation}

In this appendix, we show \cite{Fleer_book} how the recurrence
relation for $G(z,s)$ in Eq.(\ref{eq:recurrence_relation}) in
continuous form reduces to the Edwards equation
(\ref{eq:Edwards_equation_z}). First, we rewrite the recurrence
relation for $G(z,s)$ as:
\begin{equation}
G(z,s+1) \,\, e^{U_{\rm SF}(z)/k_{\rm B} T} =
G(z,s) + \lambda \left[ \, G(z-1,s) - 2 \, G(z,s) + G(z+1,s) \, \right] \,,
\label{eq:A1}
\end{equation}
where we used the expression for $G(z)$ in Eq.(\ref{eq:G(z)}). To go
from the discrete description to the continuous one, we assume that
$G(z,s)$ is a slowly varying function of $z$ and $s$. This means that:
\begin{equation}
G(z,s+1) = G(z,s) + \frac{\partial}{\partial s} \, G(z,s) + \ldots \,,
\label{eq:A2}
\end{equation}
and, keeping in mind that the distance $z$ was rescaled by $a$, that
\begin{equation}
G(z \pm 1,s) \, \longrightarrow \, G(z \pm a,s) = G(z,s) \pm a \,
\frac{\partial}{\partial z} \, G(z,s) + \frac{a^2}{2} \,
\frac{\partial^2}{\partial z^2} \, G(z,s) + \ldots \,.
\label{eq:A3}
\end{equation}
When $G(z,s)$ is a slowly varying function, it implies that the
potential $U_{\rm SF}(z)$ is close to its asymptotic value $U_{\rm
SF}(z\rightarrow\infty)\!=\!0$, so that we can expand
\begin{equation}
e^{U_{\rm SF}(z)/k_{\rm B} T} \approx 1 + \frac{U_{\rm SF}(z)}{k_{\rm
B} T} \,.
\label{eq:A4}
\end{equation}
Combining Eqs.(\ref{eq:A2})-(\ref{eq:A4}) with Eq.(\ref{eq:A1}) then
gives:
\begin{equation}
\frac{\partial}{\partial s} \, G(z,s) = \lambda \, a^2 \,
\frac{\partial^2}{\partial z^2} \, G(z,s) - \frac{U_{\rm SF}(z)}
{k_{\rm B} T} \, G(z,s) \,,
\label{eq:A5}
\end{equation}
which, for $\lambda\!=\!1/6$, is the Edwards equation in Eq.(\ref{eq:Edwards_equation_z}).

\section{Asymptotic behavior of the density profile}

In this appendix, we investigate the asymptotic behavior of the
polymer segment density profile given in
Eq.(\ref{eq:solution_dphi_m}). We write Eq.(\ref{eq:solution_dphi_m}) as:
\begin{equation}
\delta \phi(z) = \frac{\phi_b \, \xi_b^2}{\pi \, d \, R_{\rm g}} \,
\int\limits_{0}^{\infty} \!\! {\rm d}t \, \cos(x t) \,\, Q(t) \,,
\label{eq:B1}
\end{equation}
where we have defined $x\!\equiv\!z/R_{\rm G}$ and
$Q(t)\!\equiv\!f(t)/g(t)$ with
\begin{eqnarray}
f(t) &\equiv& t^2 - 1 + e^{-t^2} \,, \nonumber \\
g(t) &\equiv& \frac{\varepsilon \, t^4}{4}
+ (1 + \beta \, t^2) \,\, (t^2 - 1 + e^{-t^2}) \,.
\end{eqnarray}
The integral in Eq.(\ref{eq:B1}) can be solved \cite{Math_book}
\begin{equation}
\int\limits_{0}^{\infty} \!\! {\rm d}t \, \cos(x t) \,\, Q(t) = \pi i \,
\sum {\rm R}^{\prime} \,,
\end{equation}
where
\begin{quote}
$\sum {\rm R}^{\prime} \equiv$ sum of residues of $Q(z) \, e^{i x z}$
in the upper half plane.
\end{quote}
The poles are determined by the zero's of $g(z)$ in the complex plane
(see also \cite{Evans}).  It turns out that there are an infinite
number of poles in the upper half plane. Luckily, to describe the
asymptotic behavior of Eq.(\ref{eq:B1}), it is sufficient to locate
the pole with the lowest imaginary part. Depending on the value of
$\varepsilon$ and $\beta$, this pole is either purely
imaginary---let's denote it as $z_0\!=\! i A_0$---or it is complex and
then it comes in pairs: $z_1\!=\! \pm B_1 + i A_1$, where $A_0$,
$A_1$, and $B_1$ are positive, real numbers. Taking only these three
competing poles into account, we can determine the residues with the
result:
\begin{equation}
\sum {\rm R}^{\prime} = i \, C_0 \, e^{-A_0 \, x} + 2 i \, C_1 \,
e^{-A_1 \, x} \,\, \sin(B_1 \, x + \varphi) + \ldots \,,
\end{equation}
where
\begin{equation}
C_0 = \frac{f(z_0)}{g^{\prime}(z_0)} \,, \hspace*{20pt} C_1 = \left|
\frac{f(z_1)}{g^{\prime}(z_1)} \right| \,, \hspace*{20pt} \varphi =
\arg \left( \frac{f(z_1)}{g^{\prime}(z_1)} \right) \,,
\end{equation}
and where $z_1$ is taken to be the pole in the upper right quadrant.

As a final result, we thus have for the asymptotic density profile:
\begin{equation}
\delta \phi(z) = - \frac{2 \, \phi_b \, \xi_b^2}{d \, R_{\rm g}} \,
\left[ \, \frac{C_0}{2} \, \exp(- \frac{A_0 z}{R_{\rm g}})
+ C_1 \, \exp(- \frac{A_1 z}{R_{\rm g}}) \,\, \sin(\frac{B_1 z}{R_{\rm g}} + \varphi)
\, \right] \,.
\label{eq:asymptotic_behaviour_dphi_sum}
\end{equation}
\vskip 15pt
\noindent
{\Large\bf Acknowledgments}
\vskip 10pt
\noindent
Without the continued help and interest of Josep Bonet Avalos, this
work would not have been possible. The insightful comments of Albert
Johner are, as always, very much appreciated.

\newpage

\begin{figure}
  \centering
   \includegraphics[width=100mm]{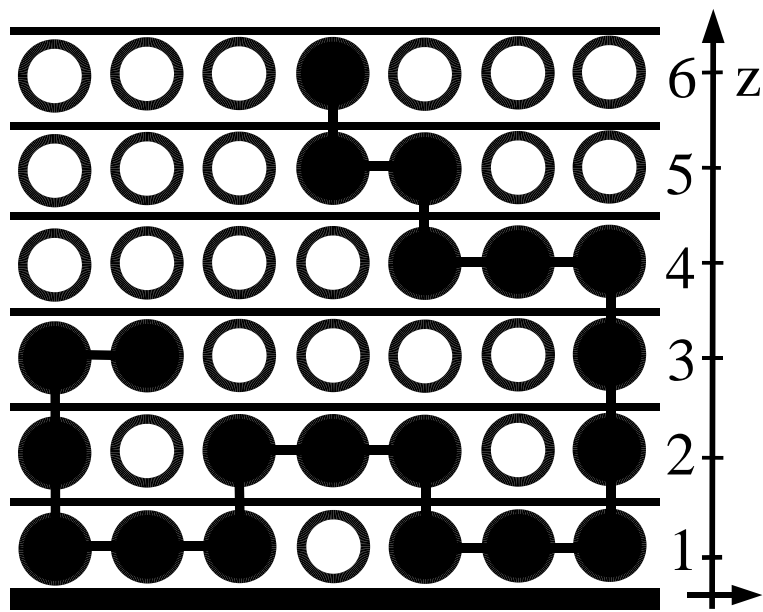}
   \caption{An adsorbed polymer on a lattice. An impenetrable wall is
located at $z\!=\!0$.}
\end{figure}

\begin{figure}
  \centering
   \subfigure[]{
   \includegraphics[width=100mm]{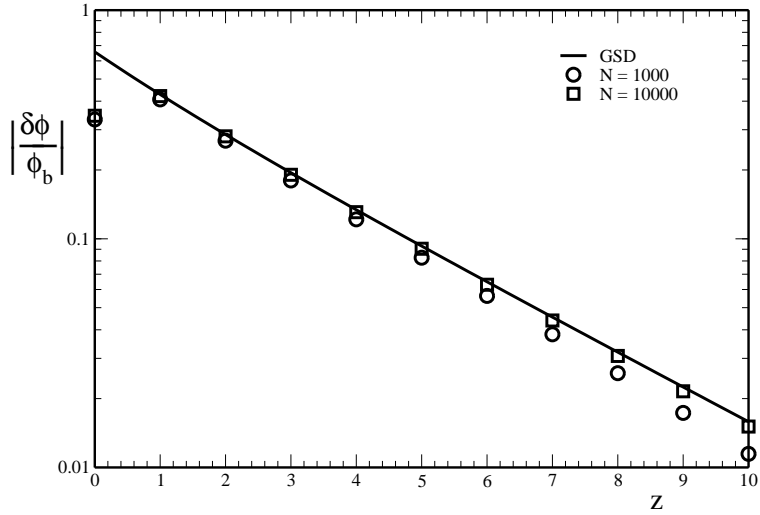}
   }
   \subfigure[]{
   \includegraphics[width=100mm]{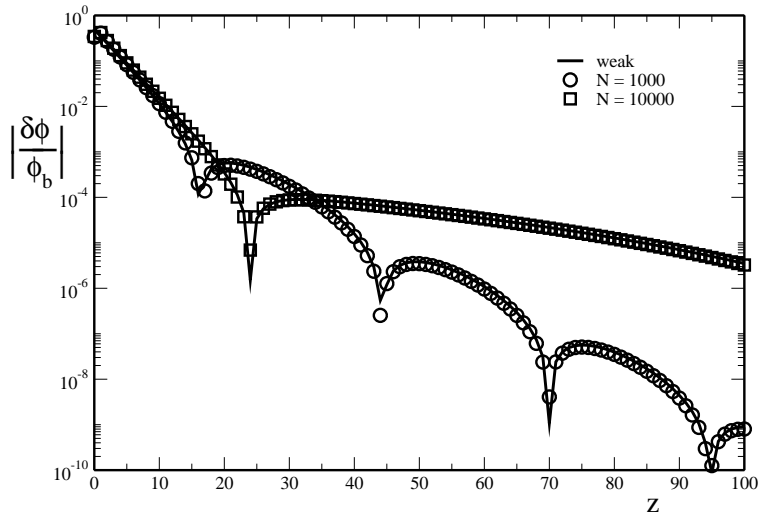}
   }\caption{Polymer segment density profiles $\delta
\phi(z)/\phi_b$ for an athermal solvent ($\chi\!=\!0$) and for
$\phi_b\!=$ 0.01.  The symbols are the SF calculations for
$N\!=\!1000$ (circles) and $N\!=\!10000$ (squares) with $\chi_{\rm
s}\!=\!0.20$. In (a) the solid line corresponds to the de Gennes
profile Eq.(\ref{eq:cotanh-profile}). In (b) the solid lines are the
density profiles calculated with Eq.(\ref{eq:solution_dphi}).}
\end{figure}

\begin{figure}
  \centering
   \includegraphics[width=100mm]{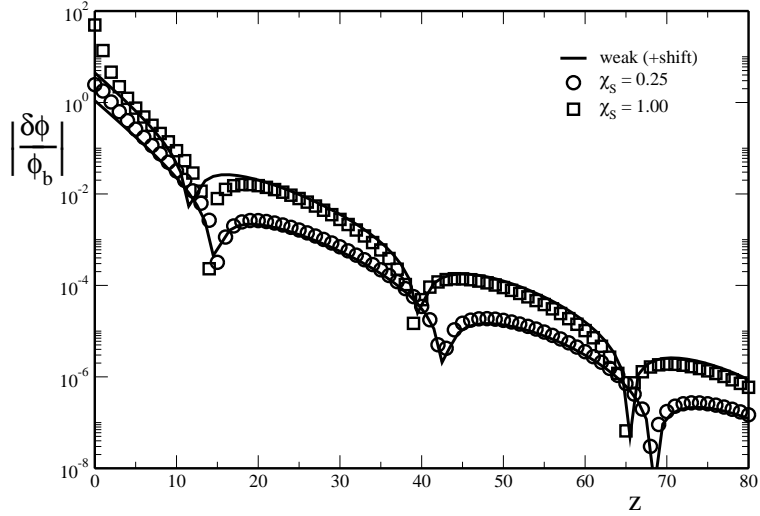}
   \caption{Polymer segment density profiles $\delta
\phi(z)/\phi_b$ for an athermal solvent ($\chi\!=\!0$) and for
$\phi_b\!=$ 0.01.  The symbols are the SF calculations for $\chi_{\rm
s}\!=$ 0.25 (circles) and $\chi_{\rm s}\!=$ 1.0 (squares) with
$N\!=\!1000$. The solid lines are the density profiles calculated with
Eq.(\ref{eq:solution_dphi}) and shifted horizontally by $-1.5 \, a$
(bottom solid line, corresponding to $\chi_{\rm s}\!=$ 0.25) and $-4.5
\, a$ (top solid line, corresponding to $\chi_{\rm s}\!=$ 1.0).}
\end{figure}

\begin{figure}
  \centering
   \includegraphics[width=100mm]{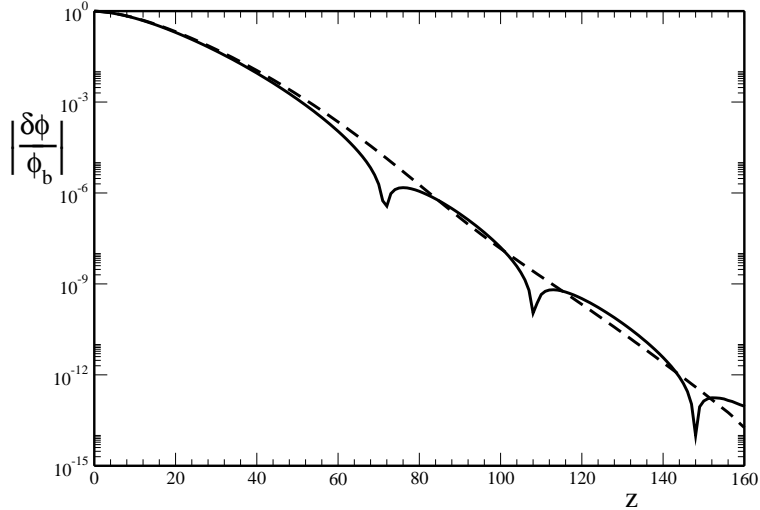}
   \caption{Polymer segment density profiles $\delta
\phi(z)/\phi_b$ for a $\Theta$-solvent ($\chi\!=\!0.5$) calculated in
the SF model; $\chi_{\rm s}\!=\!-1/12$ and $N\!=\!1000$. The bulk
density $\phi_b\!=$ 0.01 (solid line) and $\phi_b\!=$ 0.001 (dashed
line).}
\end{figure}

\begin{figure}
  \centering
   \subfigure[]{
   \includegraphics[width=100mm]{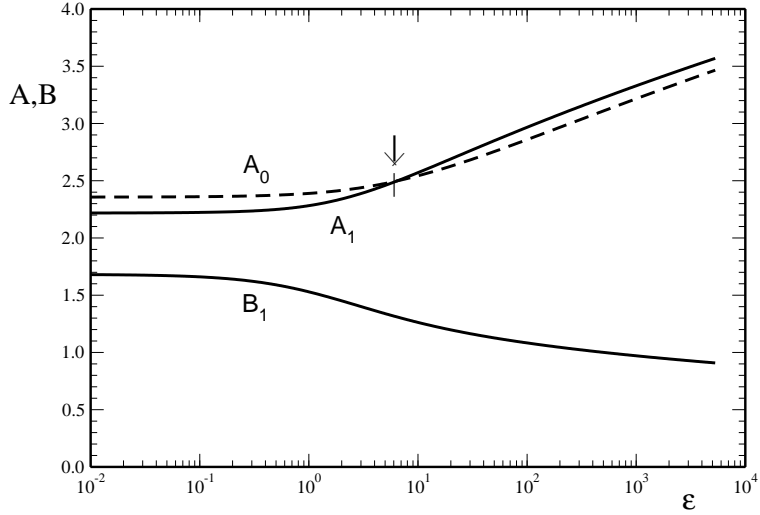}
   }
   \subfigure[]{
   \includegraphics[width=100mm]{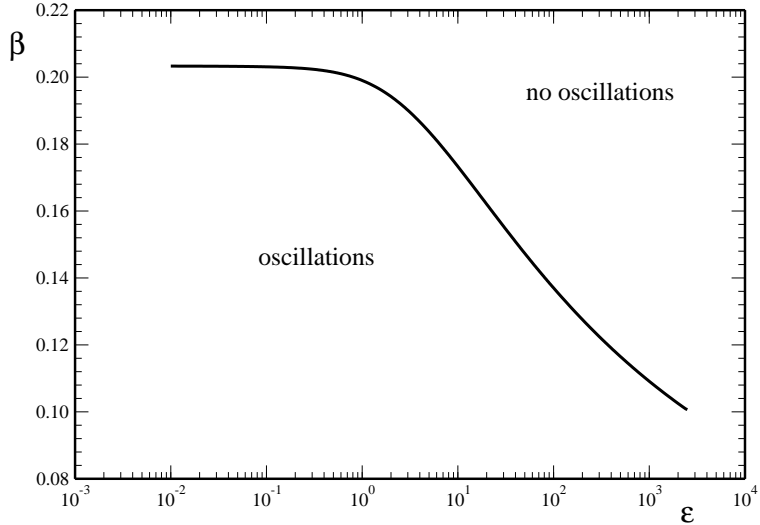}
   }
   \caption{(a) shows the coefficients $A_0$ (dashed
curve), and $A_1$, $B_1$ (solid curves), that determine the
exponential decay and the oscillation period of the polymer density
profile (see Eqs.(\ref{eq:asymptotic_behaviour_dphi_exp}) and
(\ref{eq:asymptotic_behaviour_dphi_osc})), as a function of
$\varepsilon\!=\!2/(v \phi_b N)$ for $\beta\!=\!6 m /(N a^2 v)\!=$
0.18. At $\varepsilon_{\rm FW}\!=$ 6.46398 $\ldots$, $A_0$ and $A_1$
cross. In (b) the locus of Fisher--Widom transitions is shown as a
function of $\varepsilon$ and $\beta$.}
\end{figure}

\begin{figure}
  \centering
   \subfigure[]{
   \includegraphics[width=100mm]{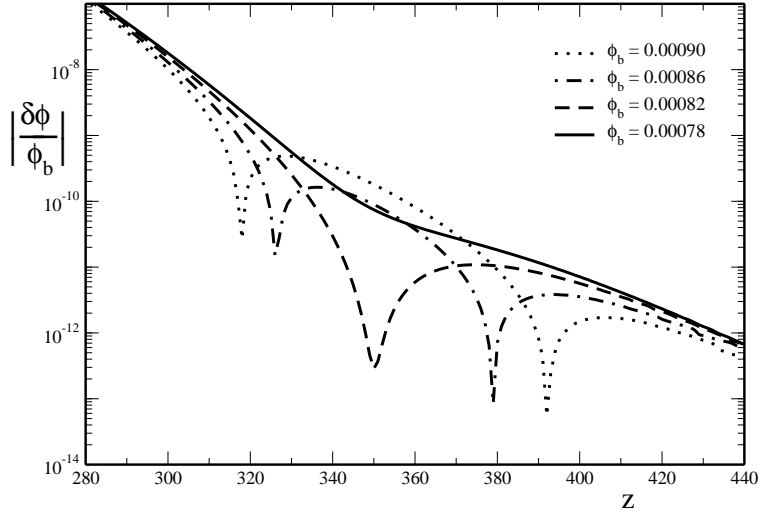}
   }
   \subfigure[]{
   \includegraphics[width=100mm]{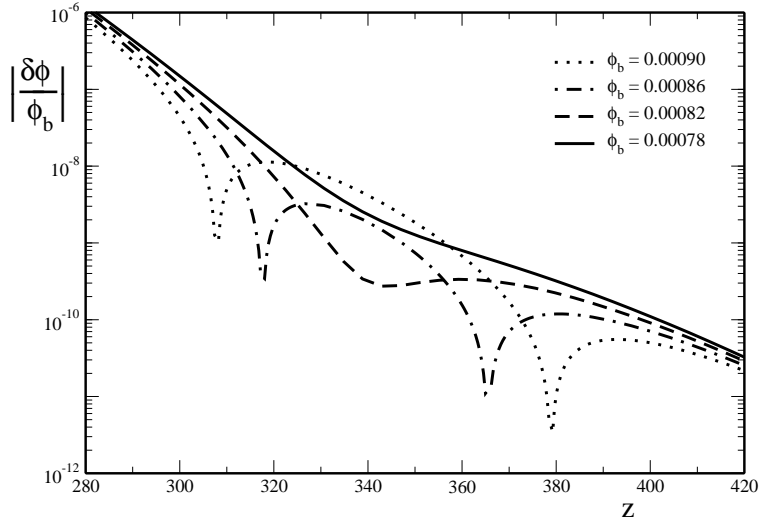}
   }
   \caption{Polymer segment density profile $\delta
\phi(z)/\phi_b$ for a $\Theta$-solvent with $N\!=\!10000$ for various
bulk densities: $\phi_b\!=$ 0.00078 (solid line), $\phi_b\!=$ 0.00082
(dashed line), $\phi_b\!=$ 0.00086 (dot-dashed line), and $\phi_b\!=$
0.00090 (dotted line). In (a), the density profiles are calculated in
the SF model, using $\chi_{\rm s}\!=\!-1/12$; in (b) the density
profiles are calculated using
Eq.(\ref{eq:asymptotic_behaviour_dphi_sum}).}
\end{figure}

\begin{figure}
  \centering
   \includegraphics[width=100mm]{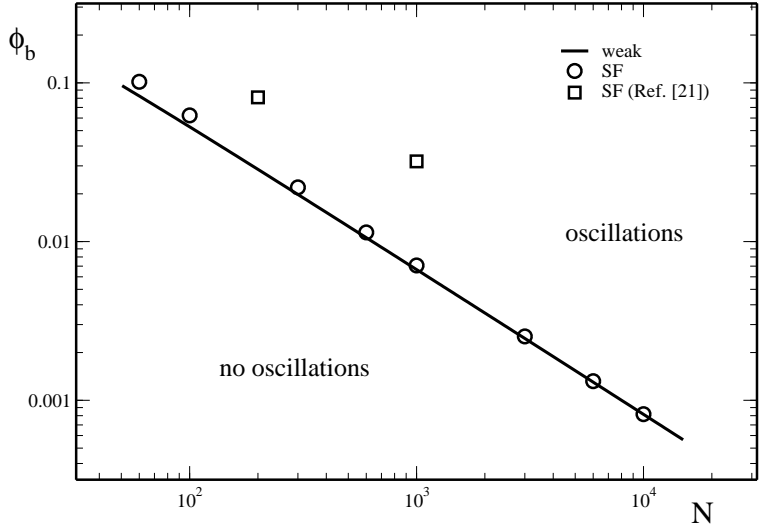}
   \caption{Bulk density corresponding to the Fisher--Widom
transition for a $\Theta$-solvent ($\chi\!=\!0.5$) as a function of
polymer chain length $N$. Symbols are the SF calculations for
$\chi_{\rm s}\!=\!-1/12$; circles correspond to the present
calculations, squares are the results reported in
Ref. \cite{Gucht}. The solid line is the analytical result (see text
for details).}
\end{figure}

\begin{figure}
  \centering
   \includegraphics[width=100mm]{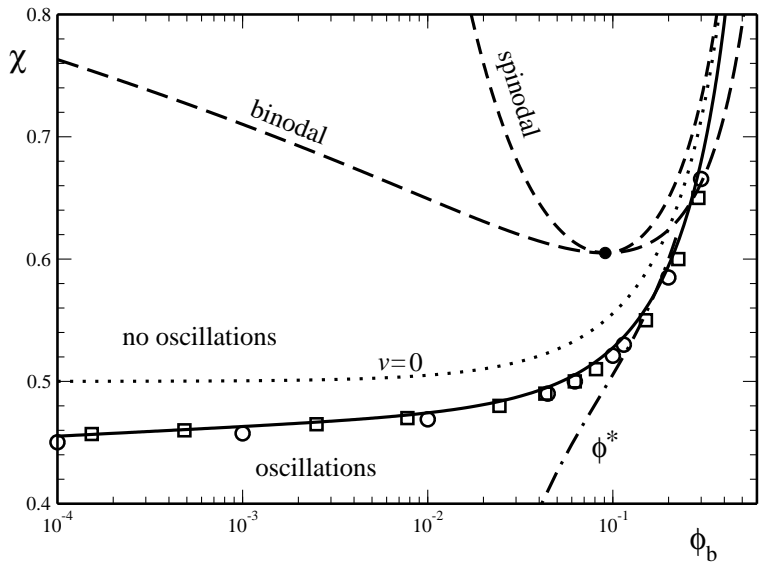}
   \caption{Fisher--Widom transition in the polymer phase
diagram for $N\!=\!100$. The symbols are the SF calculations for
$\chi_{\rm s}\!=\!0.5$ (circles) and $\chi_{\rm s}\!=\!0$
(squares). The solid line is the analytical result (see text for
details). The dotted line is where the generalized excluded volume
parameter $v\!=\!0$. The dilute to semi-dilute transition is shown as
the dot-dashed line ($\phi^* \!\approx\! 1/(v N)$). Also shown are the
Flory--Huggins spinodal and binodal regions (dashed lines).}
\end{figure}

\begin{figure}
  \centering
   \subfigure[]{
   \includegraphics[width=100mm]{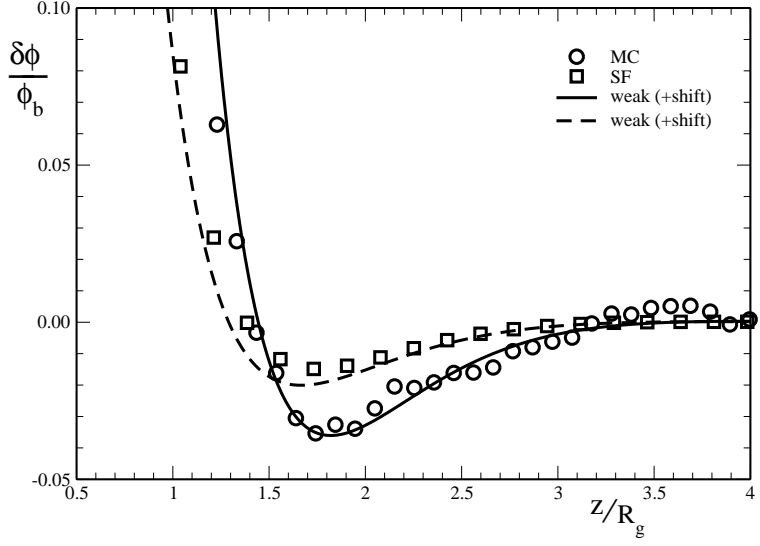}
   }
   \subfigure[]{
   \includegraphics[width=100mm]{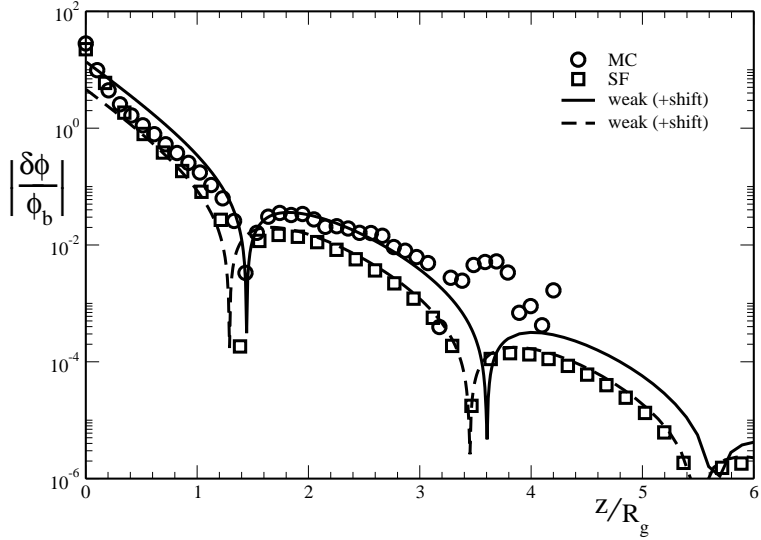}
   }
   \caption{Reduced polymer segment density profiles
($N\!=\!200$, $\phi_b\!=$ 0.0216). Circles are MC results
\cite{de_Joannis}; squares are the SF calculations for $\chi_{\rm
s}\!=\!1.0$. Solid line is the density profile calculated by
Eq.(\ref{eq:solution_dphi}) for $a/d\!=\!7.35$ and shifted by $-2 \,
a$ to fit the MC results. Dashed line is the density profile
calculated by Eq.(\ref{eq:solution_dphi}) for $a/d\!=\!4.09$
(corresponding to $\chi_{\rm s}\!=\!1.0$) and shifted by $-3.5 \, a$
to fit the SF results. (b) shows the same results on a logarithmic
scale.}
\end{figure}

\end{document}